\documentstyle[11pt,Cargesepasp,twoside,epsf]{article}
\def\edcomment#1{\iffalse\marginpar{\raggedright\sl#1\/}\else\relax\fi}
\reversemarginpar

\begin{document}
\title{Direct imaging search for planetary companions next 
to young nearby stars}
\author{Ralph Neuh\"auser}
\affil{MPI Extraterrestrische Physik, D-85740 Garching, Germany}
\author{E. Guenther}
\affil{Th\"uringer Landessternwarte Tautenburg, D-07778 Tautenburg, Germany}
\author{W. Brandner}
\affil{University of Hawaii, Institute for Astronomy, Honolulu, USA}
\author{N. Hu\'elamo, T. Ott}
\affil{MPI Extraterrestrische Physik, D-85740 Garching, Germany}
\author{J. Alves, F. Comer\'on}
\affil{European Southern Observatory, D-85748 Garching, Germany}
\author{A. Eckart}
\affil{Universit\"at K\"oln, Germany}
\author{J.-G. Cuby}
\affil{European Southern Observatory, Chile}

\begin{abstract}
We report first results from our ground-based infrared imaging 
search for sub-stellar companions (brown dwarfs and giant planets) 
of young (up to 100 Myrs) nearby (up to 75 pc) stars, 
where companions should be well separated from the central stars 
and still relatively bright due to ongoing accretion and/or contraction. 
Among our targets are all members of the TW Hya association, 
as well as other binary and single young stars either discovered 
recently among ROSAT sources (some of which as yet unpublished)
or known before. Our observations are performed mainly
with SOFI and SHARP at the ESO 3.5m NTT on La Silla
and with ISAAC at the ESO 8.2m Antu (VLT-UT1) on Cerro Paranal, 
all in the H- and K-bands. We present direct imaging data 
and H-band spectroscopy of a faint object detected next to 
TWA-7 which, if at the same age and distance as the central
star, could be an object with only a few Jupiter masses.
Our spectrum shows, though, that it is a background K-dwarf.
\end{abstract}

\section{Introduction}

Despite extensive imaging surveys, only a few sub-stellar 
companions to normal stars were detected already by
direct imaging: Gl~229~B (Nakajima et al. 1995, 
Oppenheimer et al. 1995), G196-3~B (Rebolo et al. 1998), 
and Gl~570~D (Burgasser et al. 2000),
brown dwarf companions confirmed by spectroscopy and proper motion.  
Three more candidates were presented: GG~Tau~Bb (White et al. 1999),
HR~7329~B (Lowrance et al. 2000), and CoD$-33 ^{\circ} 7795$~B
(Lowrance et al. 1999),
but spectroscopy and/or proper motion were not available so far.
Most recently, Neuh\"auser et al. (2000b) have shown that 
CoD$-33 ^{\circ} 7795$~B has spectral type M8.5 to M9 with
an optical and an infrared spectrum, both taken with the VLT,
and that it is co-moving with star A,
after two years epoch difference with $5 \sigma$ significance.
The extra-solar planet candidate directly detected by the HST near 
a T~Tauri star in Taurus (Terebey et al. 1998) has not
yet been confirmed by spectroscopy (Terebey et al. 2000).

Several extra-solar planet candidates have been detected indirectly by
radial velocity variations of stars (Latham et al. 1989, 
Mayor \& Queloz 1995, etc, see review by Marcy \& Butler 1998), 
one such candidate is confirmed 
by a transit event (Charbonneau et al. 2000).  
There is no direct imaging detection of an extra-solar planet, yet.  
Direct imaging detection of planets like those in our solar
system but orbiting other stars is difficult due to the limited
dynamical range: Extra-solar planets are simply too faint and too
close to their bright host stars. One can try to avoid the problem of
spatial resolution by searching for planetary companions around nearby
stars, where the orbit of the outermost solar system planet
corresponds to several arcsec, sufficient to resolve a faint object
next to a bright star.  However, very nearby stars usually are too
old, so that their hypothetical planets (like e.g. our Jupiter)
are correspondly too faint for direct detection with current technology.

Young planets, on the other hand, are still self-luminous due to
on-going accretion and/or contraction (Burrows et al. 1997, Brandner
et al. 1997, Malkov, Piskunov \& Zinnecker 1998) 
and, if also nearby, they would be
sufficiently bright and resolved for direct detection.  Direct imaging
of these young planets is optimal at the near-infrared bands H and K,
where the brightness difference between young stars and young planets
is expected to be the lowest (Burrows et al. 1997) and where also
the seeing is better than in the optical. In addition,
nearby stars usually have large proper motion, so that one can decide
after only a few years whether a companion candidate is co-moving.
Finally, there is a crucial advantage in studying companion planets
candidates instead of free-floating planet candidates: The mass can be
better constrained for companions than for free-floating planets,
because age and distance of the primary is usually well-known.

\section{Our sample: Young nearby stars}

We selected young ($\le 100$ Myr) nearby ($\le 75$~pc) stars, some of
them from the literature, others discovered recently among ROSAT
sources by ourselves (some of these as yet unpublished).  We are
confident on the young age of our sample of stars because of Lithium
6708\AA~absorption lines (and/or H$\alpha$ emission, IR excess and/or
kinematic membership to a young cluster). We know the distances of
most target stars from Hipparcos (and for some of them, from kinematic
membership to a cluster with known distance).  Our target list
includes all members of the TW Hya association (TWA), the T~Tauri
stars in the nearby and even younger MBM 12 cloud (Hearty et al.
2000a,b), the members of the more recently discovered Tucanae 
(Zuckerman \& Webb 2000) and HorA moving groups (Torres et al. 2000),
as well as isolated young stars like GJ 182.

\begin{figure}
\vspace{8cm}
\includegraphics{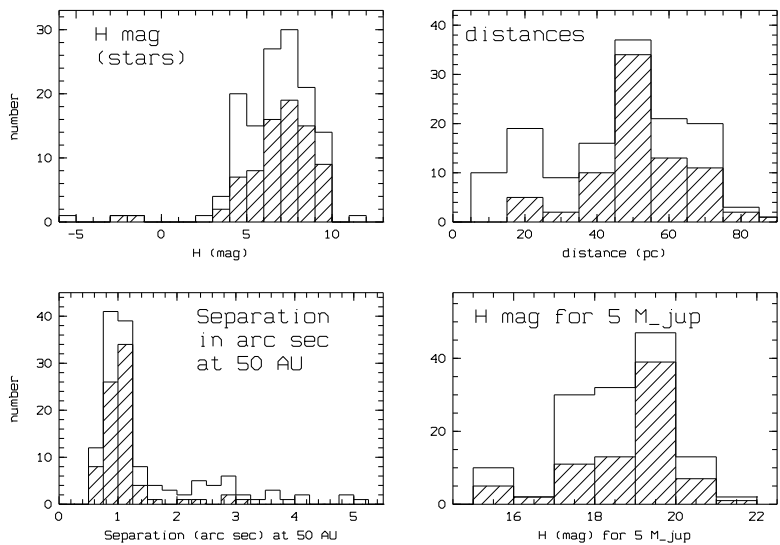}
\caption{Properties of our sample of young nearby stars and possible
companions: primary star H-band magnitude distribution (upper left),
distance distribution (upper right), distribution of angular
separations at 50 AU (lower left), and distribution of companion
H-band magnitudes (lower right), as expected from Burrows et al.
(1997) given the ages and distances of our target stars.  Hatched
are those stars recently discovered by ROSAT.}
\end{figure}

Because we know the distances and ages of the stars in our sample, we can
predict the H- and K-band magnitudes of possible substellar companions
for different masses, e.g. from 1 to $80~M_{jup}$, using the non-gray
theory by Burrows et al. (1997).  In Figure 1, we present the H-band
magnitudes of the program stars (either known from infrared
photometry, or estimated from their known spectral types and optical
magnitudes), their distance distribution, the corresponding angular
separation between those stars and possible companions at assumed 50
AU physical separations, and the K-band magnitude distribution of
assumed $\sim 10~M_{jup}$ mass companions (estimated from Burrows et
al. (1997) for the known ages and distances of the stars).  These
distributions clearly show that such companions can be detectable
and resolvable with current technology.

Hence, we have started an observational program using mainly the SOFI
and SHARP infrared cameras at the ESO-3.5m-NTT on La Silla, and the
ISAAC imaging camera at the ESO-8.2m-Antu (VLT-UT1), but also 
other state of the art instruments. 

\section{First results: Case study TWA-7}

Using the MPE-build SHARP speckle camera, we have detected a very
faint object 2.5 arcsec south-east of the young nearby star T~Tauri
TWA-7, a member of the TW Hya association (called TWA, see Webb et al. 1999).  
Four stars of the TWA association have been observed by
Hipparcos, their mean distance being 55 pc. By comparison with
isochrones, we obtain an age of 1 to 6 Myr (Neuh\"auser et al. 2000a).

The faint object TWA-7B detected by SHARP, 2.5 arc sec south-east of TWA-7A 
has $H=16.4$ and $K=16.3$ mag, which is more than 9 mag fainter than the 
primary star at these wavelengths. If TWA-7B were to be a companion to
TWA-7A, i.e. if it were to be at the same distance and age, then its
apparent H- and K-band magnitudes would correspond to absolute
magnitudes (at 55 pc) consistent with an effective temperature
T$_{eff} \simeq 1050$ K and a surface gravity $\simeq 3000$~g/s$^{2}$
(see table 5 in Burrows et al. 1997).  These values are then
consistent with an object with a mass of $\sim 3$M$_{jup}$ and an age
of $\sim 10^{6.5}$ yr (see Figure 9 in Burrows et al. 1997). This
derived age is in agreement with the ages of TWA-7A and the
other TWA members.  The angular separation of 2.5 arc sec (at a
distance of 55 pc) corresponds to a physical separation of 138 AU,
well within typical T Tauri disk sizes.

TWA-7B has previously been detected by HST Nicmos observations in the
F160W, F090M, F165M, and F180M filters.  The HST F160W image has been
shown in Neuh\"auser et al. (2000a), where a coronograph has been used
as in the F165M and F180M images.  Only in the F090M image with NIC1,
no coronograph was used. 
See Neuh\"auser et al. (2000a) for the HST magnitudes, the
position angles between TWA-7A and B, and more details on data
reduction.

Confirmation of the possibly substellar nature of TWA-7B by 
spectroscopy is required. To check for a possible spectral signature 
of a very cool object we took an H-band
spectrum of TWA-7B using ISAAC at the ESO-8.2m-Antu (VLT-UT1). In
fact, the derived temperature of $\sim 1050$ K for TWA-7B (if indeed a
companion) is similar to those of known old T-dwarfs, so that one
might, under some conditions, expect to see methane absorption
features in the spectrum of TWA-7B. The technical problem with taking
and analysing a spectrum of such a faint object so very close ($\sim$
2.5 arcsec) to a much brighter star (contrast $\sim 10^4$) is again
dynamical range.  To maximize both the separation between
TWA-7B and A and also the fraction of the light coming from TWA-7B, we
placed the slit neither along both stars (i.e.  along their Position
Angle (PA)), nor perpendicular to this PA (with only B in the slit).
In the latter case, the signal from B would be on top of the dominant
signal from TWA-7A scattered light in the collapsed spectrum (zero
separation).  In the former case (maximum separation), the light from
TWA-7A would have ``swamped'' the signal from TWA-7B.  We used,
instead, a slit orientation in between those two extremes.

\begin{figure}
\vspace{7cm}
\includegraphics{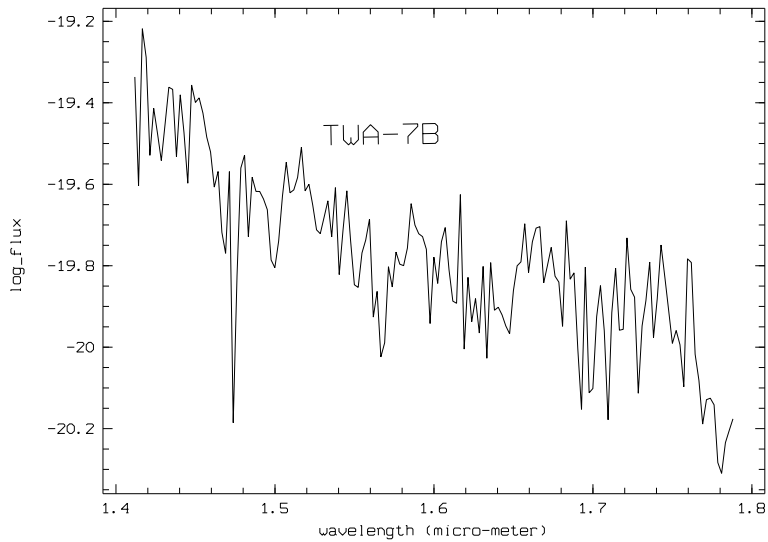}
\caption{Our ISAAC H-band spectrum of TWA-7B showing that 
it has spectral type K, i.e. that it is a background object.}
\end{figure}

We modelled the flux of TWA-7A at each
wavelength and then subtracted it from TWA-7B.
The spectrum of TWA-7B is shown in Figure 2. 
TWA-7B turned out to be of spectral type K. 
This means that, given its brightness, TWA-7B cannot be at the same 
small distance as TWA-7A (55 pc) and it is most likely a background 
K-type main sequence star.
Given that it is a faint source ($H=16.4$ mag), this K dwarf should be
at a distance between $\sim 2$ and 4 kpc, in the halo of our galaxy
($b = 21^\circ$).  Curiously, the probability of finding a 
background object as
faint as TWA-7B within a 2.5 arsec radius circle in the sky, 
towards this direction of the Galaxy, is about 1\%.
This shows how important it is to take spectra of companion
candidates.

\section{Summary}

We show that ground-based direct imaging detection of extra-solar
planets is in principle possible. As an example, we present the
detection of a very faint object ($H=16.4$, $K=16.3$ mag, i.e. $\sim
9.5$ mag fainter than a nearby young star), which could have been a
few Jupiter-mass giant extra-solar planet, given its brightness and
color, and given the fact that it is located very close to a
young star.  At the age and distance for this star, this faint
object could have been the first direct imaging detection of an
extra-solar planet. However, our H-band spectrum shows that it is
likely to be a background K-dwarf.  We demonstrate, nevertheless, 
that spectroscopic observation of such a faint object that close to a
brighter star (contrast $\sim 10^4$) is possible with current
technology from ground-based telescopes.

We are now living in very special times, where
we are starting to be able to directly image extra-solar planets
and obtain their spectra.
We believe that the first direct imaging detection of an extra-solar
planet is imminent. For achiving this goal, a large-scale survey is
essential.


\begin{thebibliography}{}

\bibitem{} Brandner W., Alcal\'a J.M., Frink S., Kunkel M., 1997, The Messenger 89, 37

\bibitem{} Burgasser A.J., Kirkpatrick J.D., Cutri R.M., et al. 2000, ApJ 531, L57

\bibitem{} Burrows A., Marley M., Hubbard W., et al. 1997, ApJ 491, 856

\bibitem{} Charbonneau D., Brown T.M., Latham D.W., Mayor M., 2000, ApJ 529, L45

\bibitem{} Hearty T., Neuh\"auser R., Stelzer B., Fern\'andez, M., Alcal\'a J.M.,
Covino E., Hambaryan V., 2000a, A\&A 353, 1044

\bibitem{} Hearty T., Fern\'andez M., Alcal\'a J.M., Covino E., Neuh\"auser R.,
2000b, A\&A, in press

\bibitem{} Latham D.W., Stefanik R.P., Mazeh T., Mayor M., Burki G., 1989, Nature 339, 38

\bibitem{} Lowrance P.J., McCarthy C., Becklin E.E., et al., 1999, ApJ 512, L69

\bibitem{} Lowrance P.J., Schneider G., Kirkpatrick J.D., et al., 2000, 
ApJ, in press

\bibitem{} Malkov O., Piskunov A., Zinnecker H., 1998, A\&A 338, 452

\bibitem{} Marcy G.W. \& Butler R.P., 1998, ARA\&A 36, 57

\bibitem{} Mayor M. \& Queloz D., Nature 378, 355

\bibitem{} Nakajima T., Oppenheimer B.R., Kulkarni S.R., et al., 1995, Nature 378, 463

\bibitem{} Neuh\"auser R., Brandner W., Eckart A., Guenther E.W., Alves J.,
Ott T., Hu\'elamo N., Fern\'andez M., 2000a, A\&A 354, L9

\bibitem{} Neuh\"auser R., Guenther E.W., Petr M.G., Brandner W., Hu\'elamo N., 
Alves J., 2000b, A\&A Letters, in press, astro-ph/0007301

\bibitem{} Oppenheimer B.R., Kulkarni S.R., Matthews K., van Kerkwijk M.H., 
1995, Science 270, 1478

\bibitem{} Rebolo R., Zapatero-Osorio M.R., Madruga S., et al., 1998, Science 282, 1309

\bibitem{} Terebey S., Van Buren D., Padgett D.L., Hancock T., Brundage M., 1998, ApJ 507, L71

\bibitem{} Terebey S., van Buren D., Matthews K., Padgett D.L., 2000, AJ 119, 2341

\bibitem{} Torres C.A.O., Da Silva L., Quast G.R., de la Reza R., Jilinski E., 2000, AJ, in press

\bibitem{} Webb R.A., Zuckerman B., Platais I., et al. 1999, ApJ 512, L63

\bibitem{} White R.J., Ghez A.M., Reid I.N., Schulz G., 1999, ApJ 520, 811

\bibitem{} Zuckerman B. \& Webb R.A., 2000, ApJ, in press

\end{thebibliography}
\end{document}